\documentclass[twocolumn,aps,prc,amsmath,amssymb]{revtex4}

\usepackage{graphicx}% Include figure files
\usepackage{wrapfig}
\usepackage{dcolumn}% Align table columns on decimal point
\usepackage{bm}% bold math
\usepackage{amsmath}

\begin{document}

\title{Large back-angle quasielastic scattering for $^{7}$Li+$^{159}$Tb} 

\author{Piyasi Biswas$^{1,3}$\footnote[2]{Present address:Shahid Matangini 
Hazra Government College for Women, Tamluk, Chakshrikrishnapur, Kulberia, 
Purba Medinipur, West Bengal - 721649, India}}
\author{A. Mukherjee$^{1,3}$}
\email{anjali.mukherjee@saha.ac.in}
\author{D. Chattopadhyay$^{1}$}
\author{Saikat Bhattacharjee$^{1,3}$}
\author{M.K. Pradhan$^{1}$\footnote[3]{Present address: Department of
Physics, Belda College, Belda, Paschim Medinipur, West Bengal - 721424, India}}
\author{Md. Moin Shaikh$^{1}$\footnote[4]{Present address: Department of
Physics, Chanchal College, Chanchal, Malda, West Bengal - 732123, India}}  
\author{Subinit Roy$^{1,3}$}
\author{A. Goswami$^{1}$\footnote[5]{Deceased}}
\author{P. Basu$^{1}$\footnote[6]{Retired}}
\author{S. Santra$^{2,3}$}
\author{S.K. Pandit$^{2,3}$}
\author{K. Mahata$^{2,3}$}
\author{A. Shrivastava$^{2,3}$}
\affiliation{
$^{1}$Saha Institute of Nuclear Physics, 1/AF Bidhannagar, Kolkata-700064, 
India\\
$^{2}$Nuclear Physics Division, Bhabha Atomic Research Centre, 
Mumbai-400085, India\\
$^{3}$Homi Bhabha National Institute, Anushaktinagar, Mumbai-400094, India
}

\date{\today}

\begin{abstract}
Quasielastic scattering excitation function at large backward angle has been
measured for the weakly bound system, $^{7}$Li+$^{159}$Tb at energies around 
the Coulomb barrier. The corresponding quasielastic barrier distribution has 
been derived from the excitation function, both including and excluding the
$\alpha$-particles produced in the reaction. The centroid of the barrier 
distribution obtained after inclusion of $\alpha$-particles was found to be 
shifted higher in energy, compared to the distribution excluding the 
$\alpha $-particles. The quasielastic data, excluding the $\alpha$-particles, 
have been analyzed in the framework of continuum discretized coupled channel 
calculations. The quasielastic barrier distribution for $^{7}$Li+$^{159}$Tb, 
has also been compared with the fusion barrier distribution for the system.   
\end{abstract}
%%\pacs{24.10.Eq, 25.70.Jj, 25.60.Pj, 25.70.Mn, 25.70.+q}

\maketitle

\section{Introduction}

          Heavy ion fusion at near-barrier energies is strongly affected 
by the internal structure of the colliding nuclei and coupling to the direct
nuclear processes, like inelastic excitation and direct nucleon transfer. 
The coupling of the relative motion to the internal degrees of freedom 
successfully explained the sub-barrier fusion enhancement observed in heavy 
ion collisions with respect to the one dimensional barrier penetration model 
calculations \cite{Beck85,Das_An}. 
          
          The coupling essentially modifies the 
effective interaction potential and in turn splits the single, uncoupled fusion
barrier into a distribution of barriers. The fusion barrier distribution, 
$D_{fus}$ for a system can be derived from the measured fusion excitation 
function as \cite{Row91}, 
\begin{equation}
D_{fus}(E)= \frac{d^{2}}{dE^{2}}\left\lfloor{E\sigma_{fus}(E)}\right\rfloor
\end{equation}
where $\sigma_{fus}(E)$ is the fusion cross section for the system at the
center-of-mass energy, E. Over the past several years of research in heavy ion 
collision,  $D_{fus}(E)$ has evolved to be a powerful tool to decipher 
the effects of coupling of various channels on sub-barrier fusion and hence 
probe the reaction dynamics of nucleus-nucleus collision \cite{Das_An}. Since 
extraction of $D_{fus}(E)$ involves second derivative of $E\sigma_{fus}(E)$, 
obtaining a meaningful barrier distribution requires very precisely measured 
fusion data.

             A similar barrier distribution can also be extracted from 
large back-angle quasielastic scattering excitation function \cite{Timmers95}. 
The quasielastic scattering is defined as the sum of all direct processes, like
elastic and inelastic scattering and transfer processes. Fusion is related to 
transmission through the barrier, whereas large back-angle quasielastic 
scattering is related to reflection at the barrier. Because of the conservation
of reaction flux, these two processes may be considered as complementary to 
each other. The quasielastic barrier distribution, $D_{qel}$ is obtained as 
\cite{Timmers95},
\begin{equation}
D_{qel}(E)= -\frac{d}{dE}\left\lfloor\frac{d\sigma_{qel}}{d\sigma_{Ruth}}\left(E\right)\right\rfloor
\end{equation}
where (${d\sigma_{qel}}/{d\sigma_{Ruth}}$) is the ratio of quasielastic 
scattering and Rutherford scattering differential cross sections at a fixed 
back-angle. As $D_{qel}$ is derived from the first derivative, unlike 
$D_{fus}$, the uncertainty associated with $D_{qel}$ is less than that 
associated with $D_{fus}$. 

             It has been observed that for heavy ion collisions 
involving tightly bound nuclei, where fusion is the most dominant reaction 
process at near-barrier energies, $D_{fus}$ and $D_{qel}$ are very similar 
\cite{Das_An,Timmers98,Hu98,Si02,Lin07}. By contrast, for very heavy systems, 
where deep inelastic processes become important, it has been argued by 
Zagrabaev \cite{Zag08} that the quasielastic barrier distribution extracted 
from the sum of elastic and inelastic backscattering processes represents the 
total reaction threshold distribution and it differs from the distribution 
derived from the fusion excitation function. For reactions, where cross 
sections of non-fusion channels are comparable to fusion cross sections, a 
deviation of $D_{qel}$ from $D_{fus}$ is expected to be seen 
\cite{Mit07,Nts07}.

              Similarly, for weakly bound systems the distribution $D_{qel}$ 
extracted only from the sum of the contribution of elastic, inelastic and 
transfer processes at large back-angle will provide information about the 
total reaction threshold distribution and not about the fusion barriers. This 
is because of the fact that for weakly bound systems, apart from other direct 
processes, breakup, transfer induced breakup and incomplete fusion (ICF) are 
very important processes competing with fusion at near-barrier energies. 
Several experimental studies of quasielastic scattering excitation function at 
large back-angle and corresponding barrier distribution have been reported for 
various systems involving weakly bound stable projectiles 
\cite{Lin07,Mon09,Mu09,Oto09,Zerv09,Zerv10,Jia10,Zerv12,Zadro13,Pal14,Moin15,
Moin20}. In most of these works, the measured quasielastic scattering 
excitation function and the corresponding $D_{qel}$ were analyzed within the 
framework of coupled reaction channel (CRC) or continuum discretized coupled 
channel (CDCC) models, while a few of these works compared $D_{qel}$ with 
$D_{fus}$. The latter works showed that for 
weakly bound systems, $D_{qel}$ is in general broader compared to $D_{fus}$. 
Also, the centroid of the distribution $D_{qel}$, with quasielastic events 
defined as the sum of elastic and inelastic scattering and transfer processes, 
is found to be shifted lower in energy than that of $D_{fus}$. But the shift 
in energy between the centroids of $D_{qel}$ and $D_{fus}$ is found to be 
different for different systems.  
For $^{6,7}$Li induced reactions with $^{64}$Ni (Z=28), though 
the peak of $D_{qel}$ is observed to be shifted towards lower energy compared 
to $D_{fus}$ for $^{6}$Li+$^{64}$Ni \cite{Moin15}, the distributions $D_{qel}$ 
and $D_{fus}$ are seen to be similar for $^{7}$Li+$^{64}$Ni \cite{Moin20}.
For $^{6,7}$Li induced reactions with $^{208}$Pb (Z=82) \cite{Lin07}, 
$D_{qel}$ and $D_{fus}$ have been reported to be similar if breakup 
contribution is included in the quasielastic scattering excitation function. 
However, for $^{6,7}$Li induced reactions with $^{197}$Au (Z=79) \cite{Pal14}, 
distributions $D_{qel}$ are seen to shift towards higher energies with respect 
to $D_{fus}$ after inclusion of breakup-$\alpha$ channel in the quasielastic 
scattering excitation functions for both $^{6}$Li and $^{7}$Li cases. The
observations reported for $^{6,7}$Li-induced reactions with $^{208}$Pb and 
$^{197}$Au are contradictory, though $^{197}$Au nucleus lies very close to 
$^{208}Pb$. In the backdrop of this scenario, it would be interesting to 
investigate the role of the structure of target nuclei while comparing 
$D_{qel}$ with $D_{fus}$ in $^{6,7}$Li-induced reactions.

    In this context, we carried out a measurement of large back-angle 
quasielastic excitation function and the corresponding barrier distribution for
the system $^{7}$Li+$^{159}$Tb, at near-barrier energies, where $^{159}$Tb is
a well-deformed target nucleus. A preliminary analysis of these measurements 
was reported in Ref. \cite{WOC}. This work is a part of our systematic 
investigation of different reaction mechanisms in $^{6,7}$Li+$^{159}$Tb 
\cite{Mukh06, Pradhan11, Pradhan13}. Complete and incomplete fusion excitation 
functions for $^{6,7}$Li+$^{159}$Tb were reported in Refs. 
\cite{Mukh06,Pradhan11}. Different processes contributing to the measured 
large $\alpha$-yield in the reaction $^{6}$Li+$^{159}$Tb were disentangled and 
reported in Ref. \cite{Pradhan13}. The primary motivation of the present work 
is to investigate the role of couplings to $^{7}$Li projectile and $^{159}$Tb 
target excitations on large back-angle quasielastic scattering process within 
the framework of coupled channel calculations. 

The present paper is organized as follows: The experimental details, along with
the results are described in Sec. II. The measured quasielastic scattering
excitation function and the corresponding barrier distribution are analyzed in 
the framework of coupled channel calculations in Sec. III. A comparison of
the $D_{qel}$ and $D_{fus}$ is discussed in Sec. IV. Finally, Sec. V summarizes 
the work.
 
\section{Experimental details}
\subsection{Experimental setup}

   The $^{7}$Li beams in the energy range 17-34 MeV, in steps of 1 MeV, from 
the 14UD BARC-TIFR Pelletron Accelerator at Mumbai, India were used to bombard
a self-supporting $^{159}$Tb target foil of thickness $\approx$1.1 mg/cm$^{2}$. 
The energies of the incident beam were corrected for the loss of energy in the 
target material at half-thickness of the target. 
To detect and identify the charged particles produced in the reaction,  
a set of four $\Delta$E-E telescopes of Si-surface barrier detectors were 
placed at $\pm$170$^{\circ}$ and $\pm$160$^{\circ}$ inside a scattering chamber
of diameter 1 m. The $\Delta$E-E telescopes were mounted at $\pm$170$^{\circ}$ 
and $\pm$160$^{\circ}$ primarily to check the 
consistency of the measured quasielastic scattering events. The thicknesses of 
the detectors of each telescope were so chosen that the charged particles lose 
part of their kinetic energies in the first detector ($\Delta$E) and stop by 
depositing the residual energies (E$_{res}$) in the second detector (E). 
However, the stop detectors used in the experiment were not thick enough to 
stop the Z=1 particles. Two Si-surface barrier detectors, of thicknesses 
300 $\mu$m and 500 $\mu$m, were placed at $\pm$20$^{\circ}$ with respect to the
beam direction for monitoring the beam and also for normalization purposes. 
In front of each telescope and monitor, a collimator was placed to define the 
solid angle.

\subsection{Data analysis and Results}
\begin{figure}[ht]
\centering
\vspace*{-1.4 cm}
\hspace*{-1.2 cm}
\includegraphics [scale=0.48]{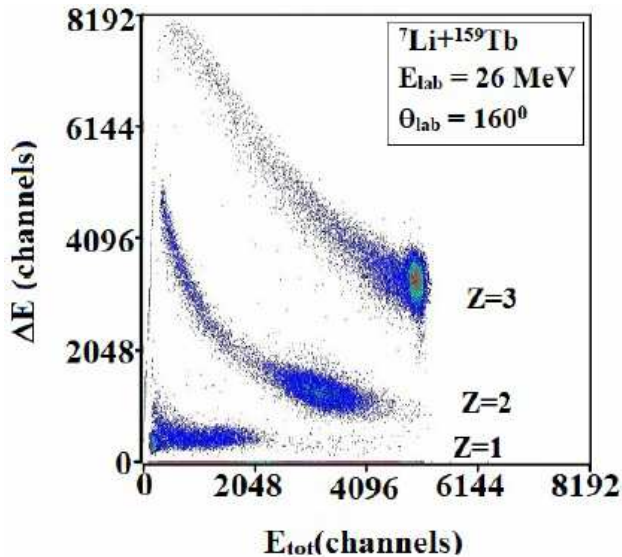}
\vspace*{-4.6 cm}
\caption{(Color online) Typical $\Delta$E-E$_{tot}$ spectrum for the 
$^{7}$Li+$^{159}$Tb reaction at E$_{lab}$=26 MeV and $\theta_{lab}$=+160$^{o}$}
\label{fig-1}
\end{figure}

        A typical two-dimensional $\Delta$E-E$_{tot}$ (where, 
E$_{tot}$=$\Delta$E+E$_{res}$) spectrum, at $E_{lab}$=26 MeV is  shown in 
Fig.~\ref{fig-1}. The peak in the Z=3
band arises from contributions due to elastic scattering of $^{7}$Li and 
inelastic scattering from the excited states of $^{159}$Tb. The low-lying 
levels of $^{159}$Tb are very closely spaced and so the inelastic excited
states of the target could not be separated from the elastic events. Besides, 
the Z=3 band 
may also contain contribution from the inelastic excitation of the projectile, 
$^{7}$Li \cite{Zadro13}. Moreover, it may also contain contribution from 
$^{6}$Li$_{g.s.}$, produced via $n$-stripping of $^{7}$Li \cite{Zadro13, 
Pandit16}, since the events corresponding to $^{6}$Li could not be separated 
from those of $^{7}$Li in the spectra. However, the Z=3 band predominantly 
consists of elastic and inelastic events. So, for nomenclature purpose, here 
we refer the Z=3 band as partial quasielastic scattering band, which 
represents primarily the sum of elastic and inelastic events. A 
one-dimensional projection of the Z=3 band was observed to show a quasielastic
peak with FWHM of $\approx$500 keV at energy, E$_{lab}$= 29 MeV. 

The Z=2 band corresponds to various events producing $\alpha$-particles in the 
reaction. A one-dimensional projection of the Z=2 band shows a broad 
$\alpha$-peak. The contribution of the $\alpha$-particles, emitted
mostly at energies corresponding to the beam velocity primarily originate from
breakup related processes, like no-capture breakup (NCBU) and ICF. The broad 
$\alpha$-peak can also have contributions from the transfer of a single 
nucleon and/ or a cluster of nucleons followed by the breakup process, and 
thereby resulting in $\alpha$-particles.
The processes that might contribute to the $\alpha$- particle cross-sections
for the $^{7}$Li+$^{159}$Tb reaction are:\\
(1) NCBU: Breakup of $^{7}$Li (B.U threshold for $\alpha + t$ is 2.47 MeV)
into $\alpha$ and $t$, either direct or sequential or both, where both the 
fragments escape without any of them being captured by the target\\
(2) Triton-ICF: $t$ captured by the target following break up of $^{7}$Li into
$\alpha$ and $t$ or a one step $t$-transfer to the target\\
(3) Neutron stripping: Single $n$-stripping from $^{7}$Li will produce $^{6}$Li
which may break into $\alpha$ and $d$, if excited above its breakup threshold
1.47 MeV\\
(4) Deuteron stripping: $d$-stripping from $^{7}$Li will produce unbound
$^{5}$He that decays to $\alpha$ and a neutron\\
(5) Proton stripping: $p$-stripping from $^{7}$Li will produce unbound $^{6}$He
which then decays to $\alpha$ and two neutrons\\
(6) Proton pickup: $p$-pickup by $^{7}$Li will lead to $^{8}$Be which
immediately decays to two $\alpha$-particles. Since this is an inclusive
measurement, each $^{8}$Be will contribute two $\alpha$-particles to the total
$\alpha$-yield. But, the contribution of $\alpha$-particles from the $p$-pickup
channel may be expected to be very small in comparison to the total 
contribution of $\alpha$-particles from other processes, like ICF 
($t$+$^{159}$Tb)
\cite{Pandit16}. Hence, the extra $\alpha$-particle contribution arising from 
the double counting of $\alpha$-particles may be neglected in comparison to the
total $\alpha$-particle contribution for the reaction.  

The Z=1 band in the figure shows a fall-back feature because the stop detectors
were not thick enough to stop the Z=1 particles. So the events corresponding to
Z=1 could not be used in the analysis. 

The ratio of quasielastic to Rutherford cross-sections is given by the
expression,
\begin{equation}
\begin{split}
\frac{d\sigma_{qel}}{d\sigma_{Ruth}}\left(E,\theta_{tel}\right)=\left\lfloor\frac{N_{qel}(E,\theta_{tel})}{N_{m}(E,\theta_{m})}\right\rfloor X\\
\left\lfloor\frac{(d\sigma_{Ruth}/d\Omega)(E,\theta_{m})}{(d\sigma_{Ruth}/d\Omega)(E,\theta_{tel})} \right\rfloor \left(\frac{\Delta\Omega_{m}}{\Delta\Omega_{tel}}\right)
\end{split}
\end{equation}
where,\\
N$_{qel}$(N$_{m}$) is the average yield in telescope (monitor) detector,\\
${\frac{d\sigma_{Ruth}}{d\Omega}}(E,\theta_{m}(\theta_{tel}))$
is the calculated Rutherford scattering cross-section at the corresponding
bombarding energy E, at monitor angle $\theta_{m}$ (telescope angle
$\theta_{tel}$), and\\
(${\frac{\Delta\Omega_{m}}{\Delta\Omega_{tel}}}$) is the solid
angle ratio of monitor to telescope detectors.\\
The ${\frac{\Delta\Omega_{m}}{\Delta\Omega_{tel}}}$ ratio for each of the four 
telescope angles was determined from the measurements at the lowest bombarding 
energies of 17, 18 and 19 MeV, where the elastic scattering is purely 
Rutherford. 

        The "partial" quasielastic counts $N_{qel}(E,\theta_{tel})$ at each 
bombarding energy were obtained from the sum of elastic and inelastic counts in
the Z=3 band in Fig.~\ref{fig-1}. As the measurements were done at 
angles close to 180$^{\circ}$, centrifugal correction was incorporated to 
obtain the effective c.m. energies ($E_{eff}$). The results of the quasielastic 
events at $\pm$170$^{\circ}$ and $\pm$160$^{\circ}$ were converted to those 
for 180$^{\circ}$ by mapping to $E_{eff}$ using the relation \cite{Timmers95},

\begin{equation}
E_{eff}=\frac{2E_{c.m.}}{1+cosec\frac{\theta_{c.m.}}{2}}
\end{equation}

To check the consistency of the data, the quasielastic excitation functions and
barrier distributions were extracted using the data taken at $\pm$170$^{\circ}$
and $\pm$160$^{\circ}$, and after appropriate centrifugal correction they were 
found to agree fairly well each other. The good agreement between the 
measurements at different angles gave us confidence in our data. 

\begin{figure}[ht]
\centering
\vspace*{-1.0cm}  
\includegraphics[width=9cm]{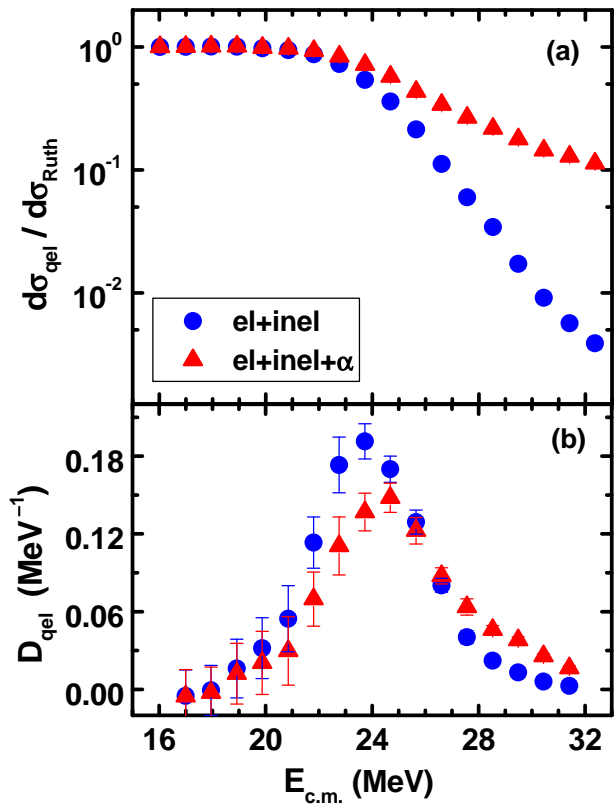}
\vspace*{-1.5cm}  
\caption{(Color online) Comparison of a) quasielastic excitation function and
b) quasielastic barrier distribution for $^{7}$Li+$^{159}$Tb, excluding
($\bullet$) and including ($\blacktriangle$) the $\alpha$ particles produced 
in the reaction.}
\label{fig-2}
\end{figure}

The "partial" quasielastic scattering excitation function determined from the 
Z=3 events and the corresponding quasielastic barrier distribution, D$_{qel}$, 
extracted using Eq. (1) are shown by the solid circles ($\bullet$) in 
Figs.~\ref{fig-2}(a) and ~\ref{fig-2}(b), respectively. The E$_{c.m.}$
energies in the figures are essentially $E_{eff}$ energies. The quasielastic 
cross sections in Fig.~\ref{fig-2}(a) were obtained by averaging the cross 
sections for the two telescopes at $\pm$170$^{\circ}$, after appropriate 
centrifugal corrections. The barrier distribution shown by the solid circles 
($\bullet$) in Fig.~\ref{fig-2}(b) was derived from the "partial" quasielastic 
excitation function, instead of the "total" quasielastic cross sections which 
would include all relevant reaction channels and not only elastic and 
inelastic events. So the derived barrier distribution, shown in 
Fig.~\ref{fig-2}(b) by the solid circles ($\bullet$), does not correspond to 
the fusion barrier distribution, but rather reflect the reaction threshold 
distribution \cite{Zag08}. The quasielastic scattering cross sections and the 
corresponding barrier distribution, determined from the sum of Z=3 (elastic + 
inelastic) and Z=2 ($\alpha$) events, are also shown in Figs.~\ref{fig-2}(a) 
and ~\ref{fig-2}(b) by the solid triangles ($\blacktriangle$).
It is observed that the inclusion of $\alpha$-particles in the definition of 
quasielastic events, shifts the centroid of the barrier distribution higher by 
$\approx$800 keV than the distribution corresponding to only Z=3 events. A 
similar observation has also been reported for the system 
$^{7}$Li+$^{197}$Au \cite{Pal14}.  

\section{Coupled Channel Calculations}
              In this section, our primary focus is to investigate the effects
of couplings between different reaction channels at near-barrier energies in 
the CRC framework, and not the relation between 
$D_{fus}$ and $D_{qel}$. So, here we considered the "partial" quasielastic 
scattering cross sections derived only from the Z=3 events. This is because 
the Z=1 particles could not be stopped in the detectors and the different 
reaction channels contributing to the Z=2 events, discussed above, could 
not be disentangled in the present inclusive measurement. The coupled channel 
calculations were carried out to analyze the measured "partial" 
quasielastic scattering excitation function and barrier distribution for 
$^{7}$Li+$^{159}$Tb, employing the code FRESCO (version FRES 2.9)
\cite{Thomp88}. 

         The primary input for the CRC calculations is the entrance channel 
optical potential which consists of Coulomb potential plus the bare nuclear 
potential. The bare nuclear potential parameters for a system are 
derived from an optical model analysis of a set of measured elastic scattering 
differential cross sections for the system. But measuring pure elastic 
scattering cross sections for the system $^{7}$Li+$^{159}$Tb is experimentally 
very difficult because of the very closely spaced low-lying excited energy 
levels of $^{159}$Tb. The only elastic scattering angular distribution data 
available in the literatute for $^{7}$Li+$^{159}$Tb are those of Ref.
\cite{Patel15}. But these data have contributions from the low-lying excited 
states of $^{159}$Tb. So, bare nuclear potential parameters could not be 
obtained for the system $^{7}$Li+$^{159}$Tb. 

\begin{figure}[ht]
\centering
\vspace*{-2.8cm}
\includegraphics[width=8.6cm]{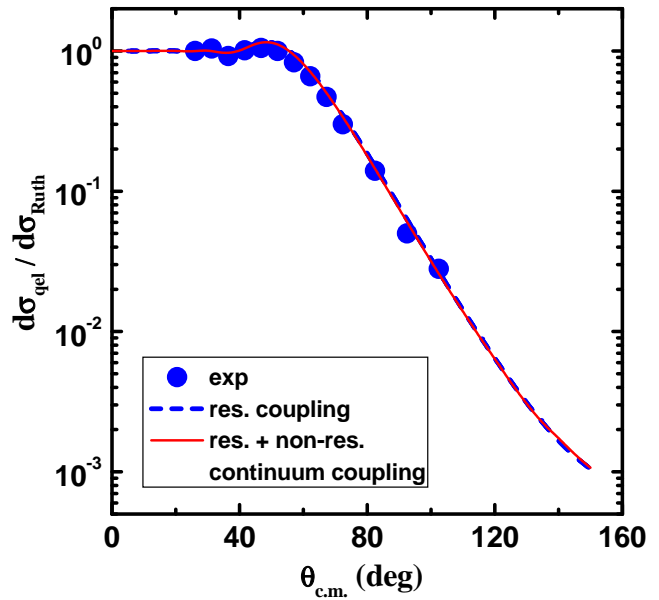}
\vspace*{-1.8cm}
\caption{(Color online) Elastic scattering angular distribution for the  
$^{7}$Li+$^{159}$Tb system at E$_{lab}$=35 MeV \cite{Patel15}. The solid and 
dashed lines represent the CDCC calculations with and without coupling to 
non-resonant continuum states of $^{7}$Li. Coupling to resonant states in the 
continuum of $^{7}$Li were included in both the calculations.}
\label{fig-3}
\end{figure}

           The elastic scattering angular distribution data \cite{Patel15} were
therefore re-analyzed in the present work, in the CDCC framework, to obtain a 
set of properly adjusted cluster folding potentials for $\alpha$+$^{159}$Tb and 
$t$+$^{159}$Tb, where the $^{7}$Li nucleus was considered to have a 
$\alpha$+$t$ cluster structure with B.U. threshold of 2.47 MeV. The CDCC 
calculations were done using the code FRESCO. The continuum of $^{7}$Li above 
the B.U. threshold of 2.47 MeV consists of non-resonant and resonant states. It
has been observed \cite{Cam16} that in the CDCC calculations for elastic 
scattering angular distribution of the relatively more weakly bound 
$^{6}$Li-induced reactions, couplings due to resonant states are more dominant 
compared to the non-resonant ones. So, for the ease of calculations, for the 
CDCC model space of $^{7}$Li we considered only the low-lying non-resonant 
continuum states up to an excitation energy of 4.4 MeV of $^{7}$Li and the two 
resonant states 7/2$^{-}$ and 5/2$^{-}$ at 4.63 and 6.68 MeV, respectively. 
Continuum states with angular momentum $l$=0, 1, 2 and 3 were considered. The 
non-resonant continuum was discretized into momentum bins of width 
$\Delta k$=0.2 fm$^{-1}$, only up to $k_{max}$=0.4 fm$^{-1}$. 
The binning of the continuum with $l$=3 was suitably done so as to include the 
two resonant states, 7/2$^{-}$ and 5/2$^{-}$ with average excitation energies 
of 2.16 and 4.21 MeV relative to the B.U. threshold of $^{7}$Li and widths of 
0.2 MeV and 3.0 MeV, respectively \cite{DC18}. The widths taken were 
sufficient enough to accommodate the main strength of the resonances. 
To calculate the bin wavefunctions, the binding potential between $\alpha$ and 
$t$ for bound and resonant states of $^{7}$Li projectile were taken from 
Ref. \cite{DC18}. The wavefunction of the projectile-target relative motion was 
expanded in partial waves up to J$_{max}$=150 and it was integrated 
numerically up to 140 fm, in steps of 0.05 fm. In addition to the continuum of 
$^{7}$Li, the bound excited state of $^{7}$Li, having spin 1/2$^{-}$ and 
E$_{ex}$=0.477 MeV, with reduced transition probability {\em{B(E2$\uparrow$)}}=
8.3 $e^{2} fm^{4}$ \cite{Pal14} was included in the coupling scheme. Also, 
the two low-lying excited states, 5/2$^{+}$ state at 0.058 MeV and 7/2$^{+}$ 
state at 0.137 MeV, of $^{159}$Tb were included in the coupling scheme 
\cite{Patel15}. The  {\em{B(E2)}} values \cite{NNDC} for the corresponding 
transitions in $^{159}$Tb, used in the calculations are listed in Table I. The 
Coulomb reduced matrix elements and the nuclear deformation lengths for the 
coupled channel calculations were derived from the {\em{B(E2)}} values, 
assuming rotational model for $^{159}$Tb. Adjusting the cluster 
folded potentials of $^{7}$Li+$^{159}$Tb, and fixing them at depth 
$V_{0}$=23.9 MeV, radius parameter, $r_{0}$=1.2 fm and diffuseness, $a$=0.5 fm 
for $\alpha$+$^{159}$Tb and at $V_{0}$=29.9 MeV, 
$r_{0}$=1.2 fm and $a$=0.5 fm for $t$+$^{159}$Tb, gave a reasonable description
of the elastic scattering angular distribution data \cite{Patel15} at incident 
energy E$_{lab}$=35 MeV. The imaginary parts of the potentials were taken to 
be of Woods-Saxon form, with $W_{0}$=50 MeV, $r_{0}$=1.0 fm and $a$=0.4 fm. 
All reorientation couplings have 
been considered in the calculations. Figure~\ref{fig-3} compares the angular 
distribution data of Ref. \cite{Patel15}, at E$_{lab}$= 35 MeV, with the CDCC 
calculated cross sections. The solid line in the figure shows the calculated 
cross sections in the above
coupling scheme, where coupling to both resonant and non-resonant parts 
of the continuum of $^{7}$Li was included. Fairly good agreement is observed 
between the calculated and measured angular distribution cross sections.
To see the importance of the effect of coupling to non-resonant part of the 
continuum of
$^{7}$Li on the elastic scattering angular distribution, the same calculations 
were repeated excluding the non-resonant continuum states of $^{7}$Li in the 
above coupling scheme. The calculated elastic 
scattering angular distribution at E$_{lab}$= 35 MeV thereby obtained are 
shown by the dashed line in Fig.~\ref{fig-3}. The very good agreement between 
the two calculations confirms that for elastic scattering angular distribution 
of $^{7}$Li+$^{159}$Tb, couplings due to resonant states of $^{7}$Li are more 
dominant compared to the non-resonant ones.

%%%%%%%%%%%%%%%%%%%%%%%%%%%%%%%%%%%%%%%%%%%%%%%%
\begin{table}
\begin{center}
\caption{Reduced transition probabilities {\em{B(E2)}} \cite{NNDC} used in the 
coupled channel analysis for the calculations of the Coulomb matrix elements 
and nuclear deformation lengths for inelastic transitions in $^{159}$Tb}
\begin{tabular}{lcc}
\\
\hline
Transition & B(E2; $J_{i}\rightarrow J_{f}$) \\
 ($J_{i}\rightarrow J_{f}$) & (e$^{2}$b$^{2}$) \\
\hline
\hline
$5/2\rightarrow3/2$& 1.87 \\
$7/2\rightarrow5/2$& 1.25 \\
$7/2\rightarrow3/2$& 0.72 \\
$9/2\rightarrow7/2$& 0.61 \\
$9/2\rightarrow5/2$& 1.13 \\
$11/2\rightarrow9/2$& 0.56 \\
$11/2\rightarrow7/2$& 1.47 \\
$13/2\rightarrow11/2$& 0.32 \\
$13/2\rightarrow9/2$& 1.74 \\
\hline
\end{tabular}
\label{Table I}
\end{center}
\end{table}
%%%%%%%%%%%%%%%%%%%%%%%%%%%%%%%%%%%%%%%%%%%%%%

\begin{figure}[ht]
\centering
\vspace{-2.2cm}
\includegraphics[width=8.6cm]{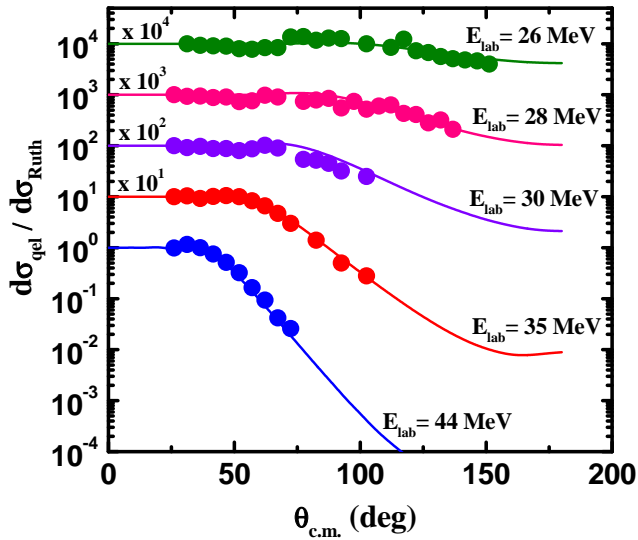}
\vspace*{-2.7cm}
\caption{(Color online) Elastic scattering angular distributions for the 
$^{7}$Li+$^{159}$Tb system \cite{Patel15}, at different bombarding energies. 
The solid lines represent the CDCC calculations.}
\label{fig-4}
\end{figure} 

Now that it is established that the effect of the resonant states of $^{7}$Li 
on the elastic scattering angular distribution of $^{7}$Li+$^{159}$Tb are 
dominant compared to the non-resonant continuum states, in the present 
re-analysis of the elastic scattering angular distribution data \cite{Patel15} 
coupling to non-resonant continuum states was not considered in further
calculations where continuum coupling is used. Hereafter, for continuum 
coupling, only the two resonant states, 7/2$^{-}$ at 4.63 MeV and 5/2$^{-}$ at 
6.68 MeV, of $^{7}$Li were considered in the CDCC coupling scheme. The CDCC 
calculations with the above coupling scheme, excluding the non-resonant 
continuum of $^{7}$Li, and using the cluster folded potentials for 
$\alpha$+$^{159}$Tb and $t$+$^{159}$Tb, as obtained above, the elastic 
scattering angular distributions for $^{7}$Li+$^{159}$Tb were calculated at 
E$_{lab}$=26, 28, 30, 35 and 44 MeV. Figure~\ref{fig-4} compares the 
angular distribution data of Ref. \cite{Patel15} with the theoretical cross
sections at different bombarding energies. The solid lines in the figure show 
the calculated cross sections. Fairly good agreement can be seen between the 
calculated and experimental elastic scattering angular distribution 
cross sections.

Having reproduced the elastic scattering angular distribution data reasonably 
well, the above coupling scheme was used to calculate the large back-angle 
quasielastic scattering excitation function for $^{7}$Li+$^{159}$Tb. The 
calculations were performed with different coupling conditions:\\
(i) No coupling to the continuum of $^{7}$Li was considered. Only inelastic 
coupling to low-lying excited states of $^{159}$Tb and bound state of $^{7}$Li 
at 0.477 MeV were included.\\
(ii) Only couplings to two resonant states, 7/2$^{-}$ and 5/2$^{-}$ at 4.63
and 6.68 MeV, in the 
continuum of $^{7}$Li were included. In this scheme, couplings to neither 
target excited states, nor bound excited state of $^{7}$Li were considered.\\
(iii) Subsequently, calculations with couplings only to the above two 
resonant states in $^{7}$Li continuum, the bound excited state of $^{7}$Li 
and the low-lying excited states of $^{159}$Tb were done.\\
(iv) Finally, full CDCC calculations with couplings to both resonant and 
non-resonant states of $^{7}$Li continuum, along with couplings to the bound 
excited state of $^{7}$Li and the low-lying excited states of $^{159}$Tb were 
performed.\\
The results of these calculations are discussed below. 
\\ 
\begin{figure}[ht]
\centering
\vspace*{-1.8cm}
\includegraphics[width=8.8cm]{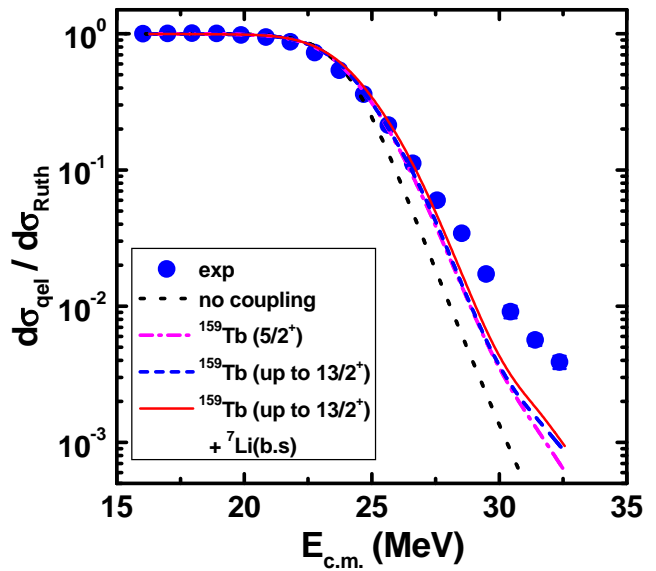}
\vspace*{-3.2cm}
\caption{(Color online) Comparison of the measured partial quasielastic 
excitation function for the $^{7}$Li+$^{159}$Tb system compared with the 
coupled channel calculations with different inelastic coupling conditions. 
See text for details.} 
\label{fig-5}
\end{figure}
\\ 
\begin{figure}[ht]
\centering
\vspace*{-2.8cm}
\includegraphics[width=8.6cm]{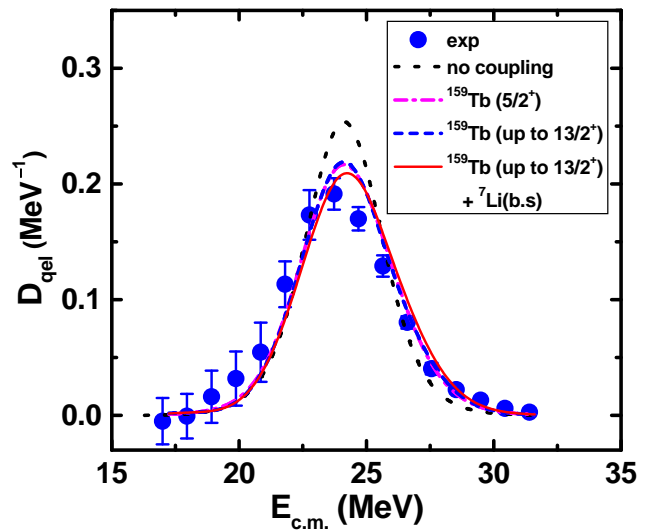}
\vspace*{-2.8cm}
\caption{(Color online) Comparison of the partial quasielastic barrier 
distribution for the $^{7}$Li+$^{159}$Tb system compared with coupled channel 
predictions with different inelastic coupling conditions. 
See text for details.} 
\label{fig-6}
\end{figure}
\\ 
The calculations were first performed with coupling condition (i).
The dot-dashed lines in Figs.~\ref{fig-5} and ~\ref{fig-6} represent the 
quasielastic scattering excitation function and the corresponding barrier 
distribution, calculated with inelastic coupling only up to the first excited 
state of $^{159}$Tb at 5/2$^{+}$. The dotted lines in the figures are the 
no-coupling calculations. It has already been mentioned in Section II that
the quasielastic peak for the Z=3 band had a FWHM of 
$\approx$500 keV at E$_{lab}$= 29 MeV. So, calculations including inelastic 
excitation of $^{159}$Tb up to 13/2$^{+}$ at 0.510 MeV in the coupling scheme, 
were done. The dashed line in the Fig.~\ref{fig-5} shows the quasielastic 
excitation function, thereby calculated. The corresponding barrier distribution
is plotted in Fig.~\ref{fig-6}. No significant change is observed either in 
excitation function or barrier distribution, if inelastic excited states of 
$^{159}$Tb beyond 5/2$^{+}$ are included in the coupling scheme. This indicates
that the 5/2$^{+}$ state of $^{159}$Tb is a strong contributor to the target 
inelastic coupling. Figures~\ref{fig-5} and ~\ref{fig-6} show that inclusion 
of only inelastic excited states of $^{159}$Tb in the coupling scheme fails to 
reproduce the quasielastic scattering excitation function and barrier 
distribution for $^{7}$Li+$^{159}$Tb. So, projectile excitation was then 
considered, in addition to the target excitation, by including the bound 
excited state of $^{7}$Li at 0.477 MeV in the coupling scheme. For comparison 
with experimental results, the quasielastic scattering cross sections were 
determined by adding the calculated elastic cross sections to the
cross sections of the inelastic states up to 13/2$^{+}$ of $^{159}$Tb and the 
bound excited state of $^{7}$Li, and are shown by the solid line in 
Fig. ~\ref{fig-5}. The corresponding barrier distribution is shown by the solid
line in Fig. ~\ref{fig-6}. The calculations are still seen to underestimate the
measured quasielastic excitation function at higher energies and also fail to 
reproduce the experimental barrier distribution. 

\begin{figure}[ht]
\centering
\vspace*{-2.5cm}
\includegraphics[width=8.6cm]{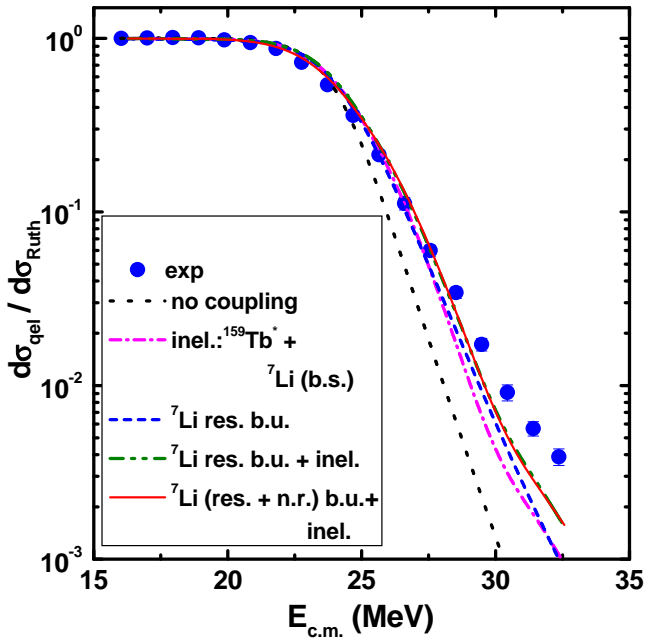}
\vspace*{-1.5cm}
\caption{(Color online) Effect of full coupling to the $^{7}$Li continuum 
in addition to the inelastic coupling to $^{7}$Li and $^{159}$Tb bound excited 
states on the quasielastic scattering excitation function for the 
$^{7}$Li+$^{159}$Tb system. See text for details.}
\label{fig-7}
\end{figure}
 
\begin{figure}[ht]
\centering
\vspace{-2.0cm}
\includegraphics[width=8.7cm]{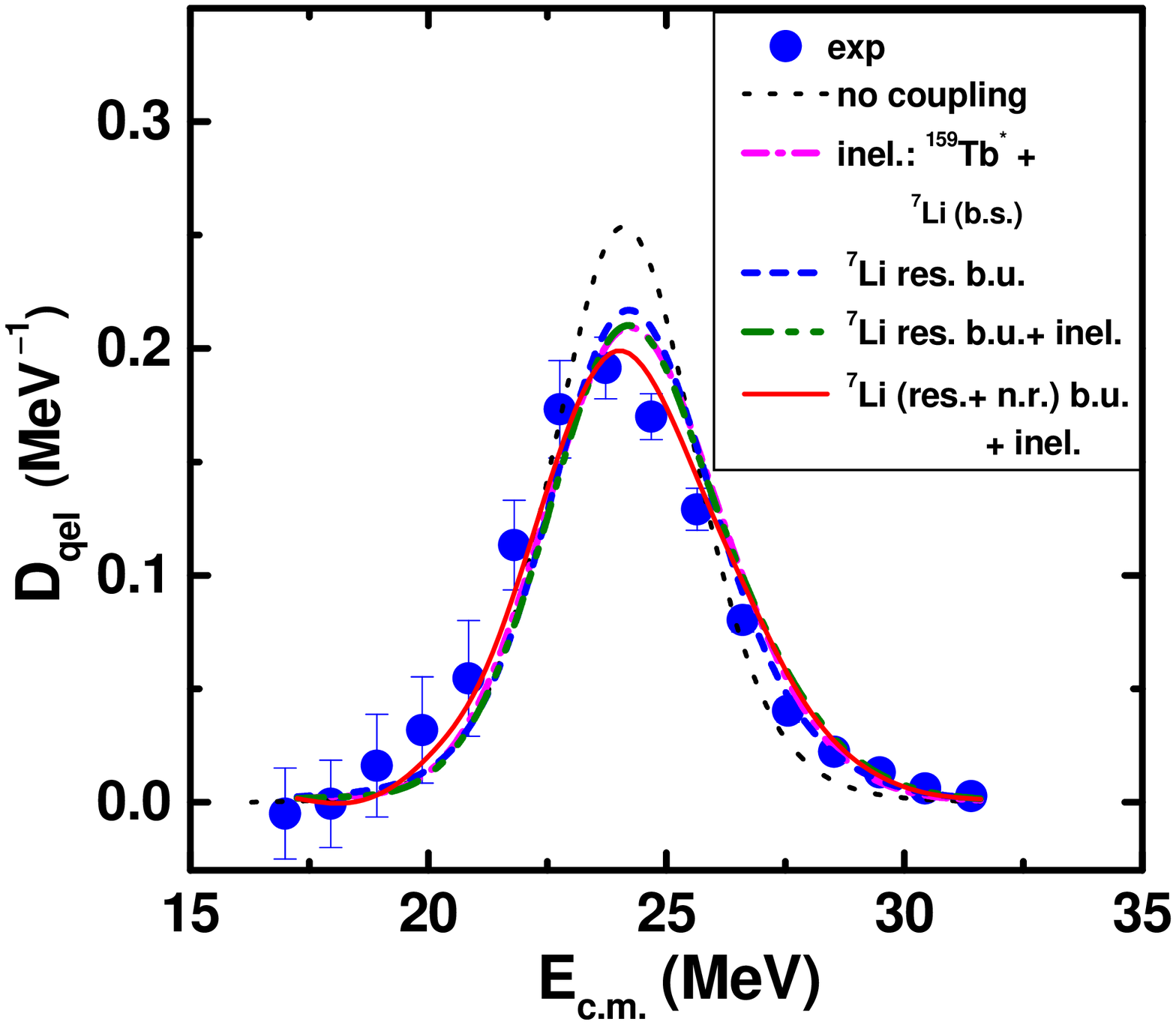}
\vspace*{-3.0cm}
\caption{(Color online) Effect of full coupling to the $^{7}$Li continuum 
in addition to the inelastic couping to $^{7}$Li and $^{159}$Tb bound excited
states on the quasielastic barrier distribution for the $^{7}$Li+$^{159}$Tb 
system. See text for details.}
\label{fig-8}
\end{figure}

The CDCC calculations were then repeated with coupling condition (ii), 
{\em{i.e.}}, couplings only to the two resonant states, $7/2^{-}$ and $5/2^{-}$
at 4.63 and 6.68 MeV, respectively, in the continuum of $^{7}$Li. The 
resulting quasielastic excitation function and barrier distribution are shown 
by the dashed lines in Figs.~\ref{fig-7} and ~\ref{fig-8}, respectively.
Subsequently, the calculations were repeated with coupling condition (iii), 
{\em{i.e.}}, couplings to the above two resonant states in the continuum of 
$^{7}$Li and also projectile and target inelastic couplings of case (i). The 
results are shown by the dot-dot-dashed lines in the figures. It can be seen 
that the inclusion of coupling to the resonant states of $^{7}$Li continuum, 
in addition to the inelastic couplings to the bound excited state of $^{7}$Li 
and the low-lying excited states of $^{159}$Tb better reproduces the 
quasielastic scattering excitation function except at higher energies, but not 
much change is observed in the barrier distribution. It is observed that the 
effect of couplings to the channels included in coupling scheme (iii) 
essentially reduces the height of the quasielastic barrier distribution and 
also broadens the distribution, as compared to the no-coupling calculations.  

Finally, full CDCC calculations were performed with coupling condition (iv), 
{\em{i.e.}}, couplings to both resonant and non-resonant states in continuum 
of $^{7}$Li along with the inelastic couplings to $^{7}$Li bound excited state 
(1/2$_{-}$, 0.477 MeV) and $^{159}$Tb low-lying excited states of case (i). The
CDCC model space of $^{7}$Li was discretized into small bins of width 
$\Delta k$=0.2 fm$^{-1}$ up to $k_{max}$=0.8 fm$^{-1}$. The resonant states were
treated appropriately to avoid double counting. Continuum states with angular 
momentum $l$=0, 1, 2 and 3 were considered. For lower bombarding energies, the 
convergence is reached by decreasing the upper limit of the excitation energy. 
Other details of the scheme are discussed above in the CDCC calculations 
for elastic scattering angular distribution.
The results of the calculations are shown by the solid lines in Figs. 7 and 
8. The inclusion of non-resonant part of the $^{7}$Li continuum in the 
calculations significantly affects the height and location of the centroid of 
the quasielastic barrier distribution, though no considerable change is
observed in the quasielastic excitation function. The height of the
barrier distribution is now almost similar to the experimental barrier 
distribution, and the centroid of the distribution has also shifted 
considerably towards the experimental barrier distribution.

It needs to be pointed out that the effects due to couplings to 
transfer and transfer induced breakup channels have not been explored here. 
Inclusion of these couplings may better reproduce the quasielastic excitation 
function at the higher energies. The small shift observed between the 
centroids of the experimental and theoretical barrier distributions could be 
due to the reaction channels, especially transfer induced breakup, not included
in the calculations.\\ 
\\
%%%%%%%%%

\section{Comparison of $D_{qel}$ and $D_{fus}$}
A comparison of the $D_{qel}$, including and excluding the $\alpha$ particle
contribution, with the $D_{fus}$ may shed some light on the importance of the
$\alpha$ particle contribution in defining the quasielastic scattering events
for weakly bound systems.

To compare $D_{qel}$ with $D_{fus}$, an attempt was made to extract the
$D_{fus}$ from the measured complete fusion (CF) excitation function for 
$^{7}$Li+$^{159}$Tb \cite{Mukh06,Broda75}. Unfortunately, $D_{fus}$ could not 
be extracted from the reported fusion cross sections \cite{Mukh06,Broda75}, 
because only a few data points were available for differentiation. Therefore, 
a rough comparison of the experimental $D_{qel}$ was made with the theoretical 
$D_{fus}$, extracted from the calculated fusion cross sections which reproduced
the measured CF cross sections \cite{Mukh06,Broda75}. 

\begin{figure}[ht]
\centering
\vspace*{-2.6cm}
\includegraphics[width=9cm]{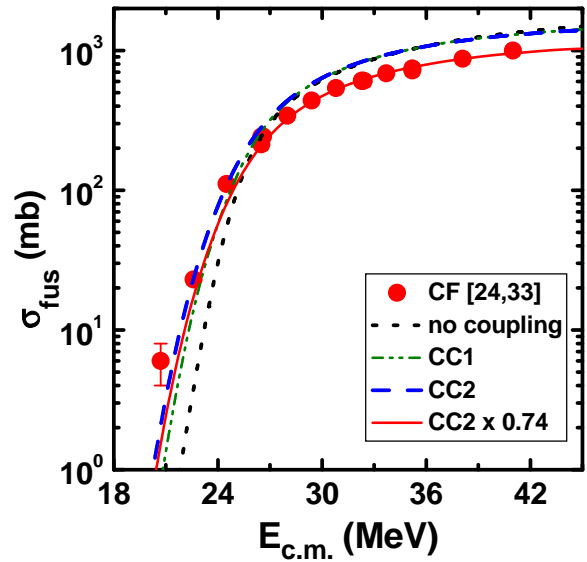}
\vspace*{-2.8cm}
\caption{(Color online) Complete fusion cross sections for the 
$^{7}$Li+$^{159}$Tb system \cite{Mukh06}. The dotted curve shows the no coupling
 calculations. The dot-dash-dashed (CC1) and dashed (CC2)lines are the coupled 
channels calculations performed with the code CCFULL \cite{CCFULL}. The solid 
line shows the coupled channels calculations (CC2) scaled by the factor 0.74.}
\label{fig-9}
\end{figure}

Following Ref. \cite{Mukh06}, the fusion cross sections were calculated using 
the coupled channels code, CCFULL \cite{CCFULL} with Akyuz Winther potential 
and all other parameters as mentioned in the reference. In addition to the 
coupling scheme used in Ref. \cite{Mukh06}, in the present work coupling to 
the bound excited state of the projectile $^{7}$Li, having spin $1/2^{-}$ 
and $E_{ex}$=0.477 MeV was also included 
using the rotational scheme \cite{Pal14}. Figure~\ref{fig-9} compares the 
experimental CF cross sections \cite{Mukh06,Broda75} with the calculated fusion
cross sections. The dotted curve shows the no coupling calculations
\cite{Mukh06}. The dot-dot-dashed line (CC1) shows the coupled channels 
calculation considering rotational coupling to six excited states of 
$^{159}$Tb \cite{Mukh06}. The dashed line (CC2) shows the coupled channels 
calculations including rotational coupling to six excited states of $^{159}$Tb 
and also rotational coupling to the bound excited state of $^{7}$Li. The 
measured complete fusion  cross sections for $^{7}$Li+$^{159}$Tb at above 
barrier 
energies are known to be suppressed by a factor of 0.74 compared to the fusion 
cross sections obtained from coupled channel calculations \cite{Mukh06}. The 
solid line in Fig.~\ref{fig-9} shows the CC2 cross sections, after being scaled
by a factor of 0.74, and will be referred hereafter as the calculated CF
excitation function. The fusion barrier distributions, $D_{fus}$ corresponding 
to calculated CC2 and CF cross sections were then obtained using equation (1).

\begin{figure}[ht]
\centering
\vspace*{-1.4cm}
\includegraphics[width=9cm]{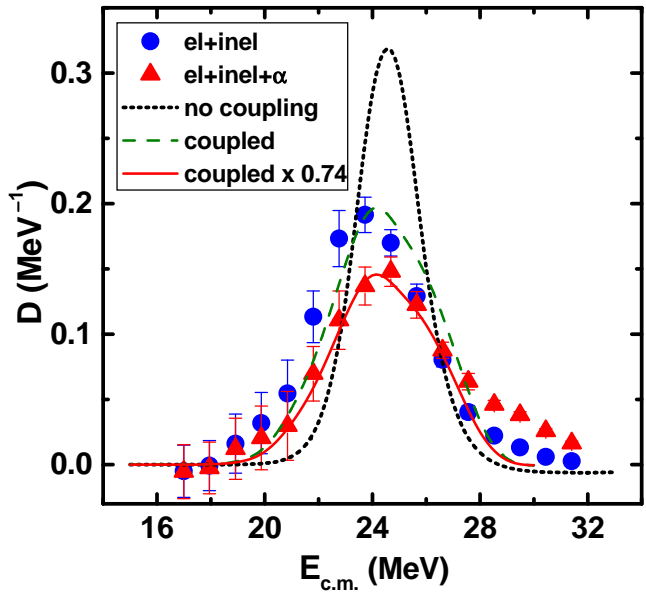}
\vspace*{-3.8cm}
\caption{(Color online) Comparison of the barrier distribution obtained from 
the quasielastic excitation function with and without inclusion of $\alpha$ 
particles for the system $^{7}$Li+$^{159}$Tb. The dotted and the dashed lines 
represent the $D_{fus}$ extracted from the coupled channel calculated  
fusion cross section without and with (CC2) coupling, respectively. The solid 
line shows the theoretical CF $D_{fus}$, obtained by scaling the CC2 cross 
sections by a factor of 0.74. The theoretical $D_{fus}$ values have been 
normalized by the factor 1/($\pi R_{b}^{2}$) to compare with $D_{qel}$. See 
text for details.}
\label{fig-10}
\end{figure}

To compare $D_{qel}$ with $D_{fus}$, the $D_{fus}$ values thus obtained were 
normalized by 1/$\pi R_{b}^{2}$, where the barrier radius $R_{b}$ was taken 
from Ref. \cite{Mukh06}. Figure~\ref{fig-10} shows a comparison of the 
experimental $D_{qel}$, including and excluding the contribution of the 
$\alpha$-particles, with the theoretical $D_{fus}$ (normalized). The dotted 
and the dashed lines represent the $D_{fus}$ extracted
from the calculated fusion cross sections without and with coupling (CC2),
respectively. The solid line shows the results obtained when $D_{fus}$
(normalized) is derived from the calculated CF cross sections \cite{Pal14}. It 
can be seen from the figure that the peak of $D_{qel}$ excluding the 
contribution of $\alpha$-particles (shown by the symbol $\bullet$) lies at an 
energy slightly lower than that of $D_{fus}$ calculated from the CC2 cross
sections and shown by the dashed line. This observation is consistent with 
those of Refs. \cite{Lin07} and \cite{Zag08}.

              The experimental $D_{qel}$ including the contribution of the 
$\alpha$-particles and shown by the symbol $\blacktriangle$ in the figure is 
found to agree reasonably well with the calculated CF $D_{fus}$, except a small
mismatch at the higher energies. Similar observation has also been reported for
the system $^{7}$Li+$^{208}$Pb \cite{Lin07}. This indicates that the agreement
of $D_{qel}$, including the contribution of $\alpha$-particles, and $D_{fus}$ 
might be independent of the structure of target nuclei. The similarity of CF 
barrier distribution with the QEL barrier distribution, including the alpha 
contribution, for $^{7}$Li-induced reactions might be understood in the 
following way.\\
For weakly bound systems, quasielastic scattering cross sections
($\sigma_{qel}$), are given by,
\begin{equation}
     \sigma_{qel}  = 1 - (\sigma_{CF} + \sigma_{ICF})
\end{equation}
where $\sigma_{CF}$ and $\sigma_{ICF}$ are CF and ICF cross sections, 
respectively. For $^{7}$Li+$^{159}$Tb reaction, $t$-capture process  is the 
dominant ICF contributor with $\alpha$-capture process being relatively 
less significant \cite{Mukh06}; an observation also reported for 
$^{7}$Li+$^{124}Sn$ \cite{Par18}. So, for $^{7}$Li-induced reactions,
$\sigma_{ICF}$ $\approx$ $\sigma_{t-capture}$. \\
It has also been reported \cite{Pandit17} that for $^{7}$Li-induced reactions,
\begin{equation}
\sigma_{t-capture}= \sigma_{\alpha} - \sigma_{\alpha-CN},
\end{equation}
where $\sigma_{\alpha}$ and $\sigma_{\alpha-CN}$ represent the inclusive 
$\alpha$-yield and the contribution of $\alpha$-particles
originating from the decay of the compound nucleus (CN) produced in the CF
process, respectively. But the CN formed in the fusion of $^{7}$Li with 
$^{159}$Tb and other heavy mass targets decays predominantly by $xn$ 
evaporation at near-barrier energies \cite{Mukh06,Pandit17}, and hence
$\sigma_{\alpha-CN}$ is expected to be negligible for such systems. Therefore, 
for $^{7}$Li-induced reactions with heavy mass targets, 
$\sigma_{ICF}$ $\approx$ $\sigma_{t-capture}$ $\approx$ $\sigma_{\alpha}$, 
which in conjunction with eqn. (5) gives\\ 
\begin{equation}
     \sigma_{qel} + \sigma_{\alpha} \approx 1 - \sigma_{CF}
\end{equation}
This perhaps explains, why $D_{qel}$ obtained from the sum of quasielastic and
$\alpha$-contributions reasonably agrees with the CF barrier distribution for
$^{7}$Li+$^{159}$Tb and $^{7}$Li+$^{208}$Pb \cite{Lin07}.

However, before reaching any conclusion, one needs to carry out simultaneous 
measurement of large back-angle quasielastic scattering and fusion excitation 
functions, and compare the corresponding experimental barrier distributions for
more $^{7}$Li-induced reactions.\\
%%%%%%%%%

\section{Summary}
The quasielastic scattering excitation function at large backward angle has 
been measured for the system $^{7}$Li+$^{159}$Tb at energies around the Coulomb
barrier and the corresponding barrier distribution has been extracted. The 
quasielastic scattering excitation function and the corresponding barrier 
distribution were determined, both with and without the contribution of 
$\alpha$-particles. The centroid of the quasielastic barrier distribution is 
seen to shift towards a higher energy with the inclusion of the contribution of
$\alpha$- particles. This corroborates the observations of Zagrabaev
\cite{Zag08} for weakly bound systems.  

The experimental "partial" quasielastic scattering cross sections and the 
barrier distribution determined only from the Z=3 events have been 
compared with the coupled channel calculations. As proper bare potential was 
not available in the literature for the system $^{7}$Li+$^{159}$Tb, the 
elastic scattering angular distribution data of Ref. \cite{Patel15} were 
re-analyzed in the CDCC framework to obtain a set of of properly adjusted 
cluster folded potentials for $\alpha$+$^{159}$Tb and $t$+$^{159}$Tb so as to 
reproduce the elastic scattering angular distribution data.

Using the cluster folded potentials for $\alpha$-$^{159}$Tb and $t$-$^{159}$Tb 
thus determined, the coupled channel calculations were done, including
different couplings at a time, in the CDCC framework to see their effects on 
the measured quasielastic scattering excitation function and corresponding 
barrier distribution separately. The inelastic coupling scheme that included 
the low-lying 
excited states of $^{159}$Tb and the bound excited state of $^{7}$Li, fails 
to reproduce the experimental quasielastic excitation function and the barrier 
distribution for $^{7}$Li+$^{159}$Tb. The quasielastic scattering excitation 
function could be reproduced reasonably well, except at the higher energies,
by including coupling to the continuum of $^{7}$Li, in addition to the above 
inelastic couplings. Though the height of the quasielastic barrier  
distribution could be reproduced reasonably well, a small shift between the 
centroids of the 
experimental and theoretical barrier distribution was observed. The shift 
might be attributed to the effects of other reaction channels, especially 
transfer and transfer induced breakup processes not considered in the work.  

As experimental $D_{fus}$ could not be obtained, the calculated fusion cross 
sections which reproduced the measured fusion cross sections of 
$^{7}$Li+$^{159}$Tb \cite{Mukh06} were used to derive the $D_{fus}$. The 
$D_{fus}$, thereby obtained was compared with the experimental $D_{qel}$, both 
including and excluding the contribution of the $\alpha$-particles. This 
comparison indicates that the distribution $D_{qel}$ including the 
contribution of $\alpha$-particles, is considerably similar to CF $D_{fus}$ 
for $^{7}$Li+$^{159}$Tb system. Similar observation was also reported for
$^{7}$Li+$^{208}$Pb \cite{Lin07}, thus showing that the similarity between
CF $D_{fus}$ and $D_{qel}$, including $\alpha$-yield, might be independent of 
the structure of target nuclei. It has been argued that this similarity of
CF barrier distribution with the barrier distribution obtained from the sum of
quasielastic and $\alpha$-particle contributions, probably lies in the
fact that the $t$-capture process is the dominant ICF process in 
$^{7}$Li-induced reactions with heavy mass targets. In order to have a better 
understanding of the role of $\alpha$-contributing channels on $D_{fus}$ and 
$D_{qel}$ in reactions with weakly bound projectiles, more simultaneous 
measurements of fusion and quasielastic barrier distributions are needed for 
$^{7}$Li-induced reactions with various target nuclei, and also for reactions 
induced by other weakly bound projectiles.\\
\\ 
     
\begin{acknowledgments}
We would like to thank the Pelletron crew for the smooth operation of the 
accelerator. Thanks are also due to Mr. Sujib Chandra Chatterjee of Saha 
Institute of Nuclear Physics for his technical support during the experiment.\\ 
\end{acknowledgments}

\end{document}